**20th Annual Conference International Research Society for Public Management**
**City University of Hong Kong, 13 - 15 April 2016**
**"Collaborative, Globalized and Interdisciplinary: Moving the Public Management Debate Forward"**

**Knowledge management metrics for Public Organizations: A literature review-based proposal**

Hector Perez Lopez-Portillo (University of Guanajuato), h.perezlopezportillo@ugto.mx
Edgar Rene Vazquez Gonzalez (University of Guanajuato), edgar10@ugto.mx
Jorge Alberto Romero Hidalgo (University of Guanajuato), jorge@ugto.mx

**Abstract:**

Knowledge Management (KM) is a relatively new phenomenon that appears in the field of Public Sector Organizations (PSO) bringing new paradigms of organizational management, challenges, risks and opportunities for its implementation, development and evaluation. KM can be seen as a systematic and deliberate effort to coordinate people, technology, organizational structures and its environment through knowledge reuse and innovation. This management approach has been established in parallel with the development and use of information and communications technologies (ICT). Nowadays more PSO are embodying KM practices in their core processes for support them, and as an advanced management strategy to create a new culture based on technology and resources efficiency. In this paper, we observed that KM can support organizational goals in PSO. The aim of this paper is to understand KM factors and its associated components, and propose KM metrics for measure KM programs in PSO. Through a critical literature review we analysed diverse studies related with KM performance indicators in PSO, then based on previous works we summarized the more convenient this purpose. We found that, in academic literature, studies about KM measurement in PSO are uncommon and emerging. As well, in the last section of this paper, we present a proposal of KM metrics for PSO, and some recommendations and practical implications for KM metrics development in PSO. This academic endeavour seeks to contribute to theoretical debate about KM measure development for KM initiatives in PSO.

1. Introduction

In this paper we review Knowledge Management (KM) measurement in Public Sector Organizations (PSO) as a core element for KM initiatives success. On the first section we review KM, PSO and KM measurement. Firstly, we analysed KM relevance for today organizations. Then we observe measurement of KM as a not enough explored issue in PSO research, then we describe PSO specific configuration, pressures, demands and goals to implement KM initiatives. Finally, on the third section we develop a systematic literature review based proposal for KM metrics in PSO. We recognized that notwithstanding of the profuse and relevant academic literature developed in this field, there is still a gap to close related to KM measurement, metrics, and tools for carry out KM measurement PSO, and the empirical evidence for evaluate performance results of KM in PSO is still limited (L.G. Pee & Kankanhalli, 2016, p. 189).

2. Knowledge Management

In our time, knowledge is a fundamental and strategic resource of organizations (Wilcox King & Zeithaml, 2003), usually related to another equally important resource: time (Ragab & Arisha, 2013, p. 873). Knowledge is the currency of the current economy (Ragab & Arisha, 2013, p. 873). And for this reason, it is perhaps the most important asset of the XXI century (Tianyong Zhang, 2010, p. 572). According to Allameh, Zamani, & Davoodi, (2011, p. 1227), knowledge encompasses mind stored ideas, facts, concepts, data and techniques of human memory. Its source is the human mind and is based on information obtained through experience, beliefs and personal values. His transformation occurs when associated with decisions or actions.

Within the Knowledge Economy (Powell & Snellman, 2004), this intangible asset has an important role in national economies to sustain economic growth and to create, gain and sustain competitive advantage (Choy Chong, Salleh, Noh Syed Ahmad, & Syed Omar Sharifuddin, 2011, p. 497; Ragab & Arisha, 2013, p. 873). In the organizational world, knowledge creation is an endless process that is also updated continuously (Nonaka & Takeuchi, 1999), and to become a knowledge based organization is an imperative for organizational today success (Bose & Ranjit, 2004).

Although Knowledge has highly recognized transformative power in organizations, it is conceptualized as an intangible resource, which by its nature is difficult to manage and measure (Dalkir, Wiseman, Shulha, & McIntyre, 2007; Massingham & Massingham, 2014, p. 225). Consequently, Knowledge Management (KM) has become increasingly necessary in organizations so that they "know what they know" (Davenport & Prusak, 1998; Kogut & Zander, 1992), and as a source of competitive advantage in rapidly changing environments (Coff, 2003). The basic aim of KM is to leverage knowledge to the organization's competitive advantage by identifying critical resources and critical areas of knowledge in order to know what organizations know and what they do well and why (Garlatti, Massaro, Dumay, & Zanin, 2014, p. 176).

Consequently, according to Jennex & Olfman (2004), successful KM materializes when knowledge is reused to improve organizational performance, from this perspective, KM is important as other organizational assets and resources, for the survival and success of the organization (Asrar-ul-Haq, Anwar, & Nisar, 2016, p. 2). Indeed, a general goal of KM is "to improve the systematic handling of knowledge and potential knowledge within the organization" (Heisig, 2009, p. 5). Thus nowadays, KM is vital not only for the success of organisations, but also for the development of societies (Ragab & Arisha, 2013, p. 877).

2.1. Knowledge Management in Public Sector Organizations

In the last years, KM research and practitioners have explored many different aspects from KM process (S. M. Allameh, Zare, & Davoodi, 2011; Chang Lee, Lee, & Kang, 2005; Ding, Liang, Tang, & van Vliet, 2014; Ho, Hsieh, & Hung, 2014; Kuah, Wong, & Wong, 2012) , systems (Chang Lee et al., 2005; Gourova & Toteva, 2014; Liao, 2003; Ragab & Arisha, 2013; Savvas & Bassiliades, 2009) , enablers (S. M. Allameh et al., 2011; Alsadhan, Zairi, & Keoy, 2008; C. S. Choy, 2006; Holsapple & Joshi, 2001; Liberona & Ruiz, 2013; Mas-Machuca & Martínez Costa, 2012; Loo Geok Pee & Kankanhalli, 2008; Pinho, Rego, & Cunha, 2012; Kuan Yew Wong & Aspinwall, 2005), benefits (Alsqour & Owoc, 2015; Edvardsson & Durst, 2013; Massingham & Massingham, 2014; Weerakkody, Irani, Lee, Osman, & Hindi, 2013; Yahyapour, Shamizanjani, & Mosakhani, 2015), risks (Gil-García & Pardo, 2005; Weerakkody et al., 2013; Yang & Maxwell, 2011) , barriers (Akhavan, Reza Zahedi, & Hosein Hosein, 2014; C. Lin, Wu, & Yen, 2012; Marouf & Khalil, 2015; Singh & Kant, 2007), outcomes (Chong Siong Choy, Yew, & Lin, 2006; H.-F. Lin, 2015; Massingham & Massingham, 2014; Migdadi, 2009) and trends (Asrar-ul-Haq et al., 2016; Tsai, 2013; E. Tsui, 2005; K. Y. Wong, Tan, Lee, & Wong, 2015).

## 2.1 Public Sector Organizations

KM in Public Sector Organizations (PSO) has raised in importance (Moffett & Walker, 2015), because it plays an important role to make them more effectives (Wiig, 2002). Actually, the KM in PSO helps governments in dealing with the challenges created by the knowledge economy (Moffett & Walker, 2015, p. 68). PSO are knowledge-intense producers and users organizations (Edge, Karen, & Edge, 2005; P. Jain, 2009; Savvas & Bassiliades, 2009).

During the last years the subject had raised its importance in the academic field (Garlatti et al., 2014; Massaro, Dumay, & Garlatti, 2015), and an increasing number of researches have been performance on this matter. But, still today, there is a lack of understanding of KM in the context of PSO (Loo Geok Pee & Kankanhalli, 2008, p. 439). KM within the PSO is an underexplored research field among the academic community, and according to Garlattir, Massaro, Dumay, & Garlatti (2015, p. 531), there is still a need to understand how KM is evolving within the context of public organizations.

Likewise, many authors claim that it is necessary to improve KM practices within PSO for the growth of developing economies (A. K. Jain & Jeppesen, 2013, p. 348). And we might not forget that PSO managers and civil servants are at the same time employees of public entities, citizens and public service users (Massaro et al., 2015, p. 176).

Certainly, several differences from KM in private sector have been pointed out by some authors (Garlatti et al., 2014; Massaro et al., 2015), mainly under the argument that PSO goals are more more difficult to measure and more conflicting than in private organizations (Titi Amayah, 2013, p. 456), or that their objectives refer to the public good (Choy Chong et al., 2011, p. 498), and that external reporting is imposed by law organizations (Garlatti et al., 2014).

Also, PSO are often considered synonymous with inefficiency and a lack of motivation to be innovative (Suwannathat, Decharin, & Somboonsavatdee, 2015, p. 528), lower motivation for the adoption of new management practices and methods (Serrano Cinca, Mar Molinero, & Bossi Queiroz, 2003), and absence of market signals as indicators *of performance* (Loo Geok Pee & Kankanhalli, 2008, p. 444).

De Angelis (2013, p. 1), argue that PSO are influenced by a growing need for competition, performance standards, monitoring, measurement, flexibility, emphasis on results, customer focus and social control (De Angelis, 2013, p. 1). Indeed in PSO, motivations for adopt KM practices respond to legislative and political changes as well as citizens' and businesses' needs (Loo Geok Pee & Kankanhalli, 2008).

Knowledge within the PSO has the potential of broadening and increasing the effectiveness of the organization's overall knowledge base (Ann Hazlett, Mcadam, & Beggs, 2008, p. 58), and to strengthen government effectiveness and competitiveness in the current changing environment (Moffett & Walker, 2015, p. 68) and to influence and improve the public sector renewal processes (Edge et al., 2005, p. 45).

Moreover, academic literature also unveils that KM in PSO should increase citizen satisfaction (Serrano Cinca et al., 2003), improve transparency (Mbhalati, 2014), reduces corruption (Elkadi, 2013; Tung & Rieck, 2005), helps to produce better knowledge (Braun & Mueller, 2014), supports public governability (Puron-Cid, 2014), contributes to society development (Ragab & Arisha, 2013), helps to take collaborative decisions (Murray E Jennex & Smolnik, 2011).

Lastly, KM in PSO is related with services cost reduction and improvement that facilitates the advancement of an information society (Management, 2004; OECD, 2003, 2015; The World Bank, 2011; United Nations, 2014), and the recover of citizens trust in PSO (United Nations, 2007). Therefore, KM is seen as an effective solution that can support public administrative activities of modernizing government (Mitre-Hernández, Mora-Soto, López-Portillo, & Lara-Alvarez, 2015). KM among PSO has been developed and extended out in parallel with the development and use of information and communications technologies (ICT) (Management, 2004; E. Tsui, 2005; United Nations, 2007).

### 2.2. Knowledge Management Measurement

As we pointed out before, regardless of the KM in PSO benefits mentioned above, the is still a lack of awareness in this sector (Chawla & Joshi, 2010). For this reason, PSO represent a unique context for practitioner and academics. From an academic perspective, KM in PSO represents and under explored research area (Massaro et al., 2015), and from a practitioner view, a background to deal with several actors (Massaro et al., 2015, p. 531) and to better managing knowledge, and to enhance partnerships with all the stakeholders for improve the overall performance of the public sector (P. Jain, 2009). Therefore, studying KM within PSO requires a separate research agenda (Massaro et al., 2015), and there is an open opportunity for researchers to develop and test specific frameworks and models for PSO (Dumay, Guthrie, & Puntillo, 2015).

For this research we analyse measurement in KM, firstly because regardless of it is a Private or Public Sector Organization, KM initiatives should demonstrate their value, and KM outcomes need to be evaluated in order asses KM implementation success. Nonetheless, developing a performance measurement model for KM is challenging in many aspects (Kuah et al., 2012, p. 9348). In fact, KM initiatives need to demonstrate their value and justify dedicated organizational resources and efforts. Without a measurable success, enthusiasm and support for KM is difficult to continue (Andone, 2009, p. 25; Chong & Chong, 2009, p. 146).

Secondly, because due its complex nature, measurement of KM is one of the least developed (Bose & Ranjit, 2004, p. 457) nor investigated issue related to KM (Chong & Chong, 2009, p. 142; Garlatti et al., 2014). Hence is highly important to establish performance measures at different stages of KM implementation even from the beginning so that its effectiveness can be identified (Chong & Chong, 2009, p. 143)

Since KM initiatives are seen as investment decisions its performance outcomes must be evaluated and measured (Tabrizi, Ebrahimi, & Delpisheh, 2011)). In fact, *m*easurement is a key element to track its the progress and effectiveness of KM initiatives (Chong & Chong, 2009; Migdadi, 2009; K. Y. Wong et al., 2015). From this perspective it is possible to perform improvement actions, based on objective judgments (Andone, 2009), associated with KM contribution to the strategic objectives of the institution (Girard & McIntyre, 2010; Vagnoni & Oppi, 2015).

According with Chong & Chong (2009, p. 146), a proper performance measurement system should be established and adopted throughout the organization, and should not be limited to measuring employees' knowledge, expertise, and individual performance. Consequently a lack of proper knowledge evaluation may lead to ignorance regarding valuable knowledge or to the duplication of redundant knowledge (Hu, Wen, & Yan, 2015, p. 1251)

In summary, KM measurement is essential in turn to ensure that its envisioned objectives are attained to organizational strategy (Bose & Ranjit, 2004) or if they need to be aligned (Massingham & Massingham, 2014). Indeed, metrics should display the areas where there are improvement opportunities in the organization (Goldoni & Oliveira, 2010, p. 302). And consequently KM measurement would improve organizational performance (Bontis, 2001). Regardless that within KM field do not have yet standardized procedures to measure knowledge within organizations (Bose & Ranjit, 2004).

### 3. The systematic literature review

The main purpose of this study is to search, identify and select relevant sources of information to obtain useful and applicable tools, techniques or instruments to measure knowledge and KM initiatives in PSO, so as to propose a literature review base metrics for KM in PSO. Accordingly, in order to accomplish the objectives of this research, the systematic literature review (SLR) was performed following the guidelines proposed by Kitchenham & Charters (2007), and the procedures developed by Ding et. al. (2014), Ali, Ali Babar, Chen, & Stol (2010), Dybå & Dingsøyr and Garlatti et al.(2014) related works. Firstly, we stablish a research protocol with inclusion and exclusion criteria in order to obtain relevant information from primary sources.

We circumscribed the SLR to a specific academic issue: the measurement of knowledge management in public sector organizations, with focus in observe all those practices carried out by academics and practitioners stored in academic databases. For that reason, the first step of this SLR was to search into the academic databases, such as Ebsco, Scopus and Web of Science, in order to obtain primary studies (from target journals, conferences, and workshops), using operators "AND" and "OR" to combine search terms.

Table 1. Search process

| Steps | Studies |
|---|---|
| 1. Studies gathered from databases using combination of search terms | 3289 |
| 2. Studies found after repeated debugging using inclusion or exclusion criteria | 1119 |
| 3. Studies reviewed based on title | 521 |
| 4. Studies reviewed based on abstract | 151 |
| 5. Studies reviewed based on introduction | 72 |
| 6. Studies reviewed based on its content | 20 |

| 7. Studies selected according with RQs and QAs | 17 |

Then according with the selection and evaluation process we chose the primary studies for this SLR. Finally, we evaluated each study aligned with RQs and Quality Attributes (QAs), previously defined, see table 2.

Table 2. Inclusion and exclusion criteria, Research questions (RQs) and Quality Attributes (QAs)

| Inclusion and exclusion criteria | (I1) Contains keywords: Knowledge Management, Public Administration, Public Sector, e-Government, Measurement, Performance evaluation, Metrics.<br>(I2): Period of publication is from 2005 to 2015.<br>(E1): The document is in another language than English, Spanish and Portuguese<br>(E2): The entire document is not longer available |
|---|---|
| **Research questions (RQs)** | RQ1: What are the main approaches or focus of KM measurement?<br>RQ2: What are main metrics, criteria of measurement proposed by literature?<br>RQ3: What are the levels of measurement proposed or implemented?<br>RQ4: What seems to be the benefits related with KM measurement in PSO? |
| **Quality attributes (QAs)** | QA1: The study is carried out in PSO<br>QA2: There is empirical validation or evidence<br>QA3: The study method is quantitative or qualitative |

**SLR Results:** Research questions
**RQ1:** Related to the main approaches for describe, analyse and perform KM measurement in PSO we observed that financial (Abdel-Maksoud, Elbanna, Mahama, & Pollanen, 2015; Chang Lee et al., 2005; Chong & Chong, 2009; Choy Chong et al., 2011; Dehghani & Ramsin, 2015; Hu et al., 2015; R. S. Kaplan & Norton, 2000; 2007; Lee & Wong, 2015; K. Y. Wong et al., 2015), quantitative (Goldoni & Oliveira, 2010; Kuah et al., 2012; Powell & Snellman, 2004) or value perspectives (Hu et al., 2015) remains popular among the academic community, and that also a mixture of perspectives, between qualitative and quantitative values (Dalkir et al., 2007; Goldoni & Oliveira, 2010; K. Y. Wong et al., 2015) is gaining attention. This could be mainly because researchers and practitioners are trying to assess knowledge and knowledge management value linking approaches in order to validate their proposals and to answer to the unique and specific PSO configuration.

Indeed, as we review before, utilitarian ends of PSO differ from private sector. Meanwhile private sector is more focused on profitably, PSO are focused on people, in how to improve the quality of public services and the decision making process to overall progress and welfare. On the other hand, non-financial perspective is focus on measuring the performance of knowledge resources (Abdel-Maksoud et al., 2015).

**RQ2:** Although yet there is no standardized metrics nor criteria for measure knowledge or KM (K. Y. Wong et al., 2015). We found that authors from selected SLR papers propose metrics and indicators that are related with the use (Ding et al., 2014) and usability of knowledge (Serrano Cinca et al., 2003) and use of KM systems (Goldoni & Oliveira, 2010; Ragab & Arisha, 2013) to produce new and different knowledge (Abdel-Maksoud et al., 2015). Moreover some studies observe indicators linked to knowledge management products (Hu et al., 2015; Powell & Snellman, 2004), and sometimes are related with the outcomes from KM process (Kuah et al., 2012), or the organizational improvements as a result of KM outcomes (Massingham & Massingham, 2014; K. Y. Wong et al., 2015) or the financial results associated or expected from KM initiatives (Chang Lee et al., 2005; Dalkir et al., 2007), for example into the Balance Scorecard (Choy Chong et al., 2011).

We confirmed that regardless that for academic literature measurement is recognized as a crucial step after the implementation of KM initiatives (Lee & Wong, 2015, p. 712), yet the boundaries and classifications of KM in PSO metrics and performances indicators are not clear nor specific. In sum, we observed that KM metrics are mainly related with use or usability of knowledge and KM products or outcomes, and also some of them are associated with KM organizational factors improvement.

**RQ3:** In relation with the levels of measurement for knowledge and knowledge management initiatives, we found that academic literature mainly evaluate KM implementation stages (Choy Chong et al., 2011), KM process performance (Chang Lee et al., 2005; Dalkir et al., 2007; Dehghani & Ramsin, 2015; Goldoni & Oliveira, 2010; Kuah et al., 2012; Lee & Wong, 2015; K. Y. Wong et al., 2015) and KM outcomes (Abdel-Maksoud et al., 2015; Bose & Ranjit, 2004; Choy Chong et al., 2011; Dalkir et al., 2007; Goldoni & Oliveira, 2010; Hu et al., 2015; Massingham & Massingham, 2014; Powell & Snellman, 2004; H. Tsui, Lee, & Lee, 2009).

**RQ4**: Finally, we agree that KM must demonstrate their value. Consequently, in relation with the potential, expected, or perceived benefits from KM measurement. We observed that these are focused on improve organizational performance (Abdel-Maksoud et al., 2015; Choy Chong et al., 2011; Kuah et al., 2012; Kuah & Wong, 2011; Lee & Wong, 2015; Ragab & Arisha, 2013; K. Y. Wong et al., 2015) increasing costumer satisfaction (Bose & Ranjit, 2004; Serrano Cinca et al., 2003) trough quality improvements, demonstrate KM initiatives value (Ragab & Arisha, 2013), long term organization benefits (Choy Chong et al., 2011), such as competitiveness, operational efficiency, innovation (Powell & Snellman, 2004), strategic alignment (Massingham & Massingham, 2014) and effectiveness or efficiency and to improve KM processes (Goldoni & Oliveira, 2010; Hu et al., 2015). And some others like improve the policy design quality (H. Tsui et al., 2009)

Lastly, limitations founded in this SLR are similar between selected studies, which suggest the fault of enough empirical demonstration or validation of academic literature proposed metrics. And also the complex process to define specific metrics for each KM initiative, and even more for KM initiatives in PSO. In deed, we agree with Massaro et al., (2015), when they argued that KM in PSO it seems to be fragmented or disarticulated.

**KM metrics based on SLR proposal**

In this section, according to performed SRL, we propose some metrics for measure and evaluate KM in PSO. Due to its complex nature, KM in PSO is different from private sector organizations, we defined metrics that could demonstrate KM initiatives value, help to continuously improve KM process, increase costumer's satisfaction or citizen's perception, and to provide sustainable long term competitive advantage for PSO. In summary, in this paper, we want to provide a comprehensive material for academic debate and advancement of KM measurement in PSO. In table 3, we present our proposal, organized by levels of analysis (knowledge, KM initiatives and KM outcomes or KM results).

Table 3. KM metrics for PSO

| Level | Metrics |
|---|---|
| Knowledge | <ul><li>Understandability, clarity, unambiguity, conciseness, retrievability, traceability, consistency, correctness, completeness, reusability, credibility (Ding et al., 2014, p. 555), and use (Serrano Cinca et al., 2003) of knowledge.</li><li>Use of KM systems (Goldoni & Oliveira, 2010; Ragab & Arisha, 2013)</li><li>Number of frequent KMS users (Kuah et al., 2012)</li><li>Amount of knowledge stored in traditional/manual filing systems and computers</li><li>Quality of knowledge stored</li><li>Number of intellectual properties owned (Lee & Wong, 2015)</li></ul> |
| KM initiatives (also, KM process) | <ul><li>Increase in the quality of KM process outcomes (Chang Lee et al., 2005)</li><li>Increase in the capability of knowledge acquisition (Lee & Wong, 2015), creation, generation, application and utilization (Lee & Wong, 2015)</li><li>Increase on contributions to organizational repositories (Goldoni & Oliveira, 2010)</li><li>Number of times employees attend training/seminars/courses to acquire knowledge.</li><li>Number of times employees acquire knowledge from the owner-manager.</li><li>Number of times employees contact customers/suppliers to acquire knowledge.</li><li>Amount of time spent browsing the Internet/World Wide Web to acquire knowledge.</li><li>Number of times employees access the company's knowledge repositories to acquire knowledge</li><li>Number of times employees work in teams to create new knowledge.</li><li>Number of times employees participate in brainstorming sessions to create new knowledge.</li><li>Amount of time spent codifying and storing knowledge in the company's knowledge repositories.</li><li>Amount of time spent updating the company's knowledge repositories.</li><li>Employees' level of willingness to contribute to the company's knowledge repositories.</li><li>Number of times employees participate in informal discussion to share knowledge. Frequency of having meeting sessions.</li><li>Frequency of employees using technological tools (e-mail, etc.) to transfer knowledge.</li><li>Investment in basic ICT (computer, Internet, intranet, etc.).</li><li>Investment in organizational infrastructure (meeting room, filing rack, etc.).</li><li>Frequency of organizational infrastructure maintenance.</li><li>Degree of alignment between KM strategy and business strategy.</li><li>Clarity of the company's KM strategy.</li><li>Employees' degree of awareness and support toward the company's KM strategy</li><li>Amount of budget allocated for KM initiatives.</li><li>Number of employees involved in KM initiatives.</li><li>Amount of time allocated for employees to perform KM activities.</li><li>Number of professional development activities organized for employees (Lee & Wong, 2015)</li></ul> |

| KM outcomes or KM results | - New knowledge, ideas, and solutions created per employee;
- New products, inventions, and services generated (Hu et al., 2015; Powell & Snellman, 2004)
- Math model, research output, scientific paper (Hu et al., 2015)
- Documents and articles accessed or downloaded per employee;
- Documents and articles uploaded or updated per employee;
- Active communities of practice, research groups, and special interest groups;
- Communications per employee per month
- Problems solved and ideas implemented per employee;
- Knowledge assets generated per year (Kuah et al., 2012)
- Acceleration of Learning times
- Increase in sharing experiences
- Decrease of capability gap between staff skills and job requirements
- Increase of search cycle efficiency or time taken to find knowledge necessary to perform a new task
- Improvement of *stakeholders* perception of the value of the organisation
- Increase of internal reward and recognition
- Improvement of staff morale and productivity (Massingham & Massingham, 2014)
- Improvement of employee capabilities;
- Efficient use of an allocated budget
- Quantity of products or services provided
- Increase of customer satisfaction / citizens perception
- Rewarding employees under his/her supervision (Abdel-Maksoud et al., 2015)
- Increase of perceptions of internal communication (Goldoni & Oliveira, 2010)
- Revenue generated from intellectual properties
- Number of new knowledge, ideas and solutions created
- Amount of rewards given to employees who create new knowledge, ideas and solutions
- Number of times employees apply useful proposals/ideas in practice
- Number of times employees apply knowledge to solve problems
- Number of new products/services launched (Lee & Wong, 2015) |
|---|---|

**Discussions and future work**

As we can see KM metrics are related and interwoven between each other. Since KM have raised in importance, many organizations have promoted KM into their core activities and process, regardless of its different natures and ends, Public Sector Organizations have adopted this *new* management approach, first as an imperative to promote innovation, and as an effort to provide better services for citizens. In our SLR, we observed that there is still a gap in KM measurement development in this field.

We also found that KM measurement in PSO is not analysed from a market value perspective, but from its knowledge quality and quantity, KM process, KM initiatives and from KM outcomes, and potential or demonstrated benefits. KM measurement seems to be observed from different perspectives (value, non-value, outcomes and results based), from diverse approaches (qualitative and quantitative, or financial and non-financial measurement), and since many levels (knowledge, KM initiatives, KM outcomes or results and KM contribution to organizational performance).

PSO will continue to experience changes in the way they delivers services, depending on the social, political and economic realities of the day (Guthrie & Dumay, 2015, p. 259). For instance, KM has the potential of broadening and increasing the effectiveness of the organization's overall knowledge base (Ann Hazlett et al., 2008, p. 58), measurement need a plan that identifies specific knowledge assets into the PSO (Dalkir et al., 2007, p. 1449), in order to provide a valid frame for evaluate KM initiatives, KM process and outcomes. Also, PSO are increasingly using information technologies to collaborate each other, that implies a greater need to develop strong capabilities in sharing, applying, and creating knowledge (L.G. Pee & Kankanhalli, 2016, pp. 188–189)

Finally, we claim for future research focus on develop more sophisticated and comprehensive measurement methods. Attending the complex nature of knowledge, integrating artificial intelligence systems, optimization techniques (Hu et al., 2015, p. 1267) big data mining analysis and attending to specific organizational PSO settings (Garlatti et al., 2014; Massaro et al., 2015).